\documentclass[aps,twocolumn,showpacs,superscriptaddress,nofootinbib]{revtex4-2}

\usepackage[utf8]{inputenc}
\usepackage[english]{babel}
\usepackage[T1]{fontenc}
\usepackage{csquotes}

\usepackage{amsmath,amsfonts}

\usepackage{graphicx}
\usepackage{float}
\usepackage{algorithm}
\usepackage{algpseudocode}
\usepackage[dvipsnames, table]{xcolor}
\definecolor{lightgray}{gray}{0.9}
\usepackage[normalem]{ulem}
\usepackage{siunitx}
\usepackage{listings}
\usepackage{hyperref}
\usepackage{cleveref}

\usepackage{qcircuit}

\newcommand{\ket}[1]{\left| #1 \right>} 
\newcommand{\bra}[1]{\left< #1 \right|} 

\newcommand{\gv}[1]{\ensuremath{\mbox{\boldmath$ #1 $}}} 

\DeclareSIUnit\angstrom{\text{\AA}}

\begin{document}

\title{Sequence of penalties method to study excited states using VQE}

\author{Rodolfo Carobene}
\email{r.carobene@campus.unimib.it}
\affiliation{Dipartimento di Fisica, Universit\`{a} di Milano-Bicocca, I-20126 Milano, Italy}

\author{Stefano Barison}
\affiliation{Institute of Physics, \'{E}cole Polytechnique F\'{e}d\'{e}rale de Lausanne (EPFL), CH-1015 Lausanne, Switzerland}

\author{Andrea Giachero}
\affiliation{Dipartimento di Fisica, Universit\`{a} di Milano-Bicocca, I-20126 Milano, Italy}
\affiliation{Bicocca Quantum Technologies (BiQuTe) Centre, I-20126 Milano, Italy}
\affiliation{INFN - Sezione di Milano Bicocca, I-20126 Milano, Italy}

\begin{abstract}
We propose an extension of the Variational Quantum Eigensolver (VQE) that leads to more accurate energy estimations and can be used to study excited states.
The method is based on the introduction of a sequence of increasing penalties in the cost function.
This approach does not require circuit modifications and thus can be applied with no additional depth cost.

Through numerical simulations, we show that we are able to produce variational states with desired physical properties, such as total spin and charge.
We assess its performance both on classical simulators and on currently available quantum devices, calculating the potential energy curves of small molecular systems in different physical configurations.
Finally, we compare our method to the original VQE and to another extension, obtaining a better agreement with exact simulations for both energy and targeted physical quantities.
\end{abstract}

\maketitle

\section{Introduction}
\label{sec:introduction}

\begin{figure*}[!ht]
    \centering
    \includegraphics[width=1\textwidth]{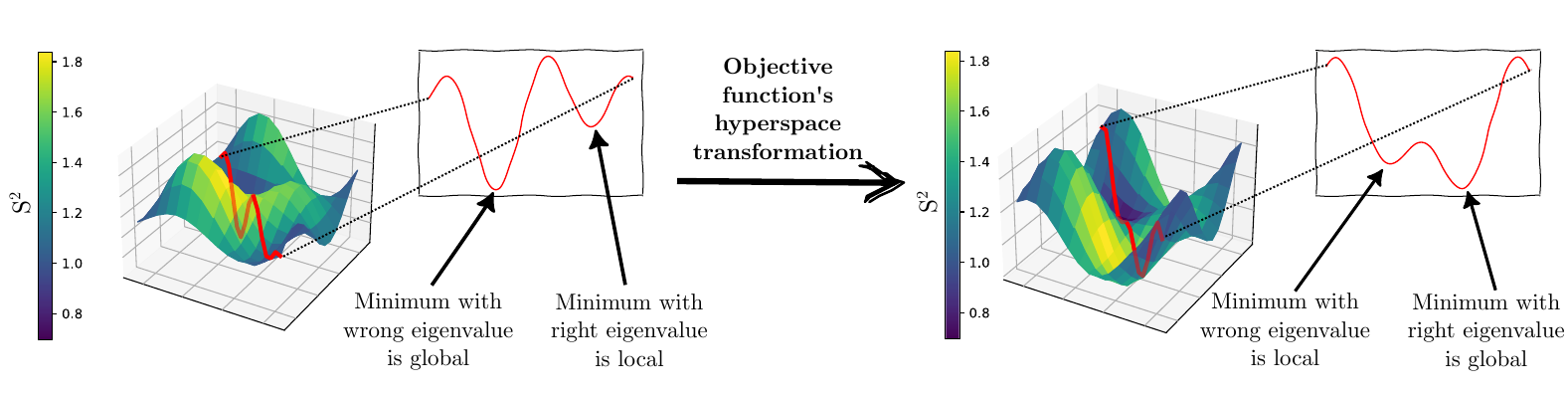}
    \caption{ Sketch of the cost function space of $\mathrm{H_3^+}$ projected on one and two dimensions.
    A comparison of the minima before and after applying a penalty constrain on the total spin shows that the desired state becomes the global minimum after the transformation.
    }
    \label{fig_hyperspace}
\end{figure*}

In recent years, the development of quantum computing hardware made strides from the first working prototypes to devices with more than a hundred qubits \cite{Ball2021}.
Companies such as IBM and Google have revealed plans to build thousands or even a million qubits devices by the end of the decade. 
However, it has to be highlighted that the number of qubits is not the only metric to take into account in a quantum computing platform and the current amount of operations that can be performed is strongly limited by hardware noise and decoherence.
While quantum error correction schemes have been proposed and even tested on hardware \cite{schindler_2011,Kelly2015,anderson_2021,Egan2021,Krinner_2022}, their large-scale application remains still far in the future.

In order to provide meaningful results in the near-term, hybrid algorithms that combine quantum devices with classical computers have been proposed \cite{Bharti2022, Beckey2020, Cerezo2020, Motta2019, Ollitrault2020, Farhi2014, Cao2019, Li2017,Barison_2021, Avkhadiev2019, Cervia2021}.
In these schemes a quantum computer is only in charge of a subroutine, while the classical computer governs the whole algorithm.
This greatly reduces the number of quantum operations required and those methods have been demonstrated to be naturally robust with respect to certain hardware errors \cite{McClean_2016}.

One such scheme is the Variational Quantum Eigensolver (VQE) \cite{Peruzzo2014, O'Malley2016}, where a parameterized quantum circuit is used to approximate the ground state of an interacting quantum system.
In this case a quantum computer is used to measure the energy and its derivatives, then a classical optimizer will tune the variational parameters according to those.
VQE has been widely studied and extended to improve the final result precision or to lower the hardware requirements \cite{Tilly2021, Colless2018, Kandala2017, Tang2021}.

More recently, extensions of VQE to the study of excited states have been proposed.
As an example, in the quantum subspace expansion method \cite{takeshita2020increasing}, excitation operators are applied to the variational ground-state, while the quantum imaginary time evolution algorithm \cite{Motta2019} can be used to construct the Lanczos subspace \cite{Yeter-Aydeniz2020}.
Alternatively, one can directly compute excitation energies using the quantum equation of motion method \cite{Ollitrault2020}, or even use algorithms of quantum machine learning \cite{Tilly2020}.
Finally, another category of VQE variants is based on the idea of modifying the cost function to guide the optimization to the desired state \cite{Nakanishi2019, Ryabinkin2019, Higgott2019, Jones2019, Danbo2020, Garcia-Saez2018}. 

In this paper, we expand on the last approach and propose a method based on the introduction of a sequence of increasing penalties in the cost function that leads to better approximated eigenstates and can be used to study excited states.
We chose to name our algorithm Sequence of Penalties VQE (SPVQE).
Based on a cost function modification, the proposed method does not need more resources than VQE.

The structure of this paper is as follows: in \Cref{sec:methods} we review the VQE algorithm, its constrained modification (CQVE) and present the Sequence of Penalties VQE method, while in \Cref{sec:results} we apply it to the study of excited and ionized state of small chemical compounds, assessing the performances both on classical simulators and on real hardware.
Finally, \Cref{sec:discussion} concludes the paper with some considerations and outlooks on the proposed method.

\section{Methods}
\label{sec:methods}

\subsection{VQE: Variational Quantum Eigensolver}
\label{sec:vqe}

Here we briefly review the Variational Quantum Eigensolver (VQE), for a more detailed explanation we suggest to refer to \cite{Peruzzo2014, McClean_2016}.

Consider a physical system described by the Hamiltonian $H$.
The aim of the VQE is to prepare a variational ground state approximation  of this system on a quantum computer.
First, the Hamiltonian is mapped into a Pauli operator \cite{Wigner1993,Seeley2012, Bravyi2002}, in order to be able to measure expectation values on quantum hardware. 
Then, the qubits are prepared in the state $\ket{\psi(\gv \theta_0)} = V(\gv \theta_0) \ket{0}$, where $V(\gv \theta)$ is a combination of parameterized quantum operations (gates) depending on the set of parameters $\gv \theta$, and $\gv \theta_0$ is the initial choice of parameters.

We use the quantum device to measure the expectation value of $H$, defined as $E(\gv \theta)=\bra{\psi(\gv \theta)}\hat H \ket{\psi(\gv \theta)}$, and its derivatives with respect to the parameters $\gv \theta$ \cite{schuld2018der,mari2020estimating,parrish2019hybrid,crooks2019gradients}.
Finally, we feed these quantities to a classical optimizer that determines new values of $\gv \theta$ in order to decrease $E(\gv \theta)$.
When this iterated procedure converges, we obtain a set of parameters $\gv \theta^{\star}$ that defines the desired state.

Now suppose that, given some operator $\hat A$, we want to prepare the eigenstate corresponding to its eigenvalue $a$.
Given that these configurations may not correspond to the ground state of the system, we can not rely solely on lowering the energy.

More generally, we may be interested in studying an excited state in a region in which quantum states with different physical properties have similar energies. 
In all these cases, VQE may fail to correctly approximate the desired state.

In perspective of using VQE to study molecules with complex energy spectra, we must develop a robust method that preserves every physical property we may desire to fix.

\subsection{CVQE: Constrained Variational Quantum Eigensolver}
\label{sec:cvqe}

Some of the proposed variations of VQE already include in the optimization process information about additional operators \cite{Moll2016, Bravyi2017, Barkoutsos2018}. 
In particular, we will briefly recall the Constrained Variational Quantum Eigensolver \cite{Ryabinkin2019}.

While in the standard VQE approach we minimize the energy as stated in \Cref{sec:vqe}, the CVQE method introduces a redefined cost function $F(\gv \theta)$ by adding a penalty multiplier for each of the operators we may desire to fix:

\begin{multline}
    \label{eq:simplepenalty}
    F(\gv \theta, \gv \mu) = \bra{\psi (\gv \theta )} \hat H \ket{\psi (\gv \theta )} + \\ + \sum_i \mu_i \left[\bra{\psi (\gv \theta )} \hat A_i \ket{\psi (\gv \theta )} - a_i \right]^2
\end{multline}

here $\hat A_i$ with $i \in \{1,\dots, m \}$ is a set of operators, $\mu_i$ are their corresponding multipliers and $a_i$ the eigenvalues that we want to fix.
Chosen $\mu_i$ sufficiently large, the desired state will correspond to a global minimum of the cost function.
In \Cref{fig_hyperspace} we show as an example the variational landscape computed for the $\mathrm{H_3^+}$ molecule, obtained by scanning the Hamiltonian and the spin operator on two random parameters.
We see that the desired state is transformed from a local to a global minimum through the application of the constrain.
We highlight that, among all the possible states satisfying the given constraints, the global minima will be always the lowest energy state.
For this reason, we envision the combination of our method with other algorithms, such as the Quantum Subspace Expansion \cite{takeshita2020increasing} or the quantum Equation of Motion \cite{Ollitrault2020}, to provide even more accurate estimations of excited state properties of physical systems.

In a practical usage of CVQE, however, the combination of an unknown good starting point $\gv \theta_0$ and restricted local knowledge of the loss landscape often leads to convergence into local minima.
Moreover, a careful fine tuning of $\mu_i$ is often required since multipliers too large yield to narrow global minima, while multipliers too small do not highlight sufficiently the global minima with respect to local ones.

\subsection{SPVQE: Sequence of Penalties Variational Quantum Eigensolver}
\label{sec:spvqe}

We propose a simple, yet effective, variation of CVQE: the Sequence of Penalty VQE (SPVQE). 

We want to have a penalty large enough to exclude any local minima, while avoiding changing the cost function space too rapidly. 
Therefore, we propose to repeat the constrained optimization increasing penalty multipliers by steps \cite{Nocedal2006}.

For clarity of exposition, we will constrain a single operator.
The method can be extended to multiple constraints modifying the penalty term.

First, we choose the maximum value of the penalty multiplier ($\mu_{max}$) and the number of steps $N_s$.
If the gap between the desired state energy $E$ and the ground state one ($E_{GS}$) is approximately known, we can choose the value of $\mu_{max}$ so that it satisfies the inequality \cite{Kuroiwa2021}:
\begin{equation}\label{ineq_mumax}
    \mu_{max} \ge \frac{E_{GS} - E}{\left(\bra{\psi_{GS}} \hat A \ket{\psi_{GS}} - a\right)^2} \, .
\end{equation}

It is possible to obtain an estimation of the gap by using classical approximated methods.
Even if this classical estimation does not reach chemical accuracy, it provides an idea of the magnitude of the gap.
In general, $N_s$ and $\mu_{max}$ are hyperparameters of the SPVQE method, therefore there is not a rigorous framework to exactly set the parameters and we have to rely on heuristics.

\Cref{ineq_mumax} provides a lower limit for the maximum multiplier, but we do not have an upper limit for the single step multiplier and therefore there is no guarantee that the choice made will leada to the feasible region.
In a theoretical perspective, $\mu_{max}$ can be chosen arbitrarily large since $N_s$ compensates for the narrowing of the global minimum, hence the problem is reduced to the choice of $N_s$.

In practice, we can first set $N_s$ by considering how much time we can afford on the quantum platform, given that computational time grows linearly with $N_s$.
In our examples a number of steps of $\approx 10$ was always sufficient, as we will show in \Cref{Fig:iterations}.

We run $N_s$ instances of VQE constrained with increasing penalties. 
At each iteration $k \in \{1,\dots, N_s\}$, we compute the penalty using $\mu=\frac{\mu_{max}*k}{N_s}$, using the optimal set of parameters $\gv \theta^{\star}_{k-1}$ of the previous step as starting point.
Finally, the best loss computed over all the iterations is returned as the result.

The complete method is schematically presented in \Cref{alg_sequencepenalty}.

\begin{algorithm}[H]
\caption{Sequence of Penalties VQE}
\label{alg_sequencepenalty}
    \begin{algorithmic}[1]
        \State Given a random starting point $\gv \theta_0$ and a Hamiltonian $\hat H$
        \State Chosen a $\mu_{max}$ and a number of iteration $N_s$
        \State We define $s=\mu_{max} / N_s$
        \State Given, initially: $\mu = 0$
            \vspace{0.3cm}
        \For{ $k = 1 ,\, 2,\,3,\, \cdots, \, N_s$}
            \State Penalty term: $P = \mu \left( \bra{\psi (\gv \theta)}\hat A \ket{\psi (\gv \theta)}- a \right)^2$
            \State Define cost function: $F = \bra{\psi (\gv \theta)} \hat H \ket{\psi (\gv \theta)} + P$
            \State Start VQE with parameters $\gv \theta^{\star}_{k-1}$
            \State Save $\gv \theta^{\star}_k$
            \State Update the multiplier: $\mu \leftarrow \mu + s$
        \EndFor
            
        \Return Energy correspondent to minimum cost function $\left(E = F -  P\right)$
    \end{algorithmic}
\end{algorithm}

\subsection{Numerical simulations}
\label{sec:num_sim}
To demonstrate a practical SPVQE application, we considered molecular systems where the constrained operators were the particle number or the total spin operator:

\begin{equation}\label{eq:operators_fixed}
    \hat{A}\in \left\{ \hat S^2 ,\, \hat N \right\}
\end{equation}

We chose  a hardware efficient ansatz composed by a layer of single qubit rotations $R_y$, each one with an independent parameter, followed by an entangling layer of CNOTs.
This structure is repeated $D$ times and, at the end of the circuit, another layer of $R_y$ rotations is applied. 
As an example, for 4 qubits the circuit reads

\vspace{0.5cm}
\centerline{
    \Qcircuit @C=1em @R=0.5em {
    & & & \ustick{\times D} \\
    & \qw  & \gate{R_y} & \qw & \ctrl{1} &  \qw       & \qw       & \qw  & \gate{R_y} & \qw \\
    & \qw  & \gate{R_y} & \qw & \targ    &  \ctrl{1}  & \qw       & \qw  & \gate{R_y} & \qw \\
    & \qw  & \gate{R_y} & \qw & \qw      &  \targ     & \ctrl{1}  & \qw  & \gate{R_y} & \qw \\
    & \qw  & \gate{R_y} & \qw & \qw      &  \qw       & \targ     & \qw  & \gate{R_y} & \qw 
    \gategroup{2}{3}{3}{7}{.7em}{^\}} 
    }
}
\vspace{0.5cm}

In the following experiments we set $D=3$, unless where explicitly stated.
To limit the number of qubits required, we considered a minimal Slater-type orbital basis set constructed with 6 primitive Gaussian orbitals (STO-6G) \cite{Kandala2017, O'Malley2016}.
The qubit mapping of the Hamiltonian is done with parity mapping \cite{Seeley2012} with the two qubit reduction.
For noiseless simulations we used the SciPy Conjugate Gradient (CG) optimizer \cite{Scipy}; whereas for noisy simulations and hardware calculations we used the Nakanishi-Fujii-Todo (NFT) optimizer \cite{Nakanishi2020} that proved to give the best results.

Some final measurements were performed after the VQE procedure to improve results in noisy simulations and hardware calculations.
These are presented in \ref{app_post_processing}.
The simulations and quantum hardware computations that can be found in \Cref{sec:results} and in the appendices are performed using IBM's open source library \textit{Qiskit} \cite{Qiskit}.

\section{Results}
\label{sec:results}

In this Section we will use the SPVQE algorithm to find ground and excited states of molecular systems and compare the results to VQE and CVQE calculations.
In particular, we will consider the Born–Oppenheimer approximation of molecular Hamiltonians, that in second quantization have the form
\begin{equation}\label{eq:hamiltonian}
    \hat H = \sum_{ij} h_{ij} \hat a_i^\dagger \hat a_j + \frac{1}{2} \sum_{ijkl} g_{ijkl} \hat a_i^\dagger \hat a_k^\dagger \hat a_l \hat a_j  + E_0 \, ,
\end{equation}

where $\hat a_i^\dagger$ and $\hat a_i$ are fermionic creation and annihilation operators corresponding each to a different spin-orbital and the coefficients $h_{ij}$ and $g_{ijkl}$ are coefficient accounting for electron kinetic energy, interactions electrons-nuclei and electron-electron repulsion.
$E_0$ is the shift obtained by fixing the atomic positions and calculating their repulsion energy.

Details on the systems studied can be found in \ref{app_details}, while complete results including nuclear repulsion energy, frozen core energy and parameters obtained at every step are available on \cite{Github}.

\subsection{Using SPVQE to study excited states}

First, to assess the performance of the SPVQE algorithm, we studied the energies of the first ionic states of $\mathrm{H_2}$ defined by different particles numbers, namely $\mathrm{H_2^+}$ ($N=1$), $\mathrm{H_2^-}$ ($N=3$) and $\mathrm{H_2^{--}}$ ($N=4$).
We analyzed the energy of the molecules in different atomic configurations, varying the bond length from $0.3\,\si\angstrom$ to $3.5\,\si\angstrom$.

Calculations were performed on a classical computer simulating a noiseless quantum environment.
The circuit describing $\mathrm{H_2}$ and its excited states needs two qubits, for a total of 8 parameters when a  $D=3$ circuit is considered.

\begin{figure}[!ht]
    \includegraphics[width=1\columnwidth]{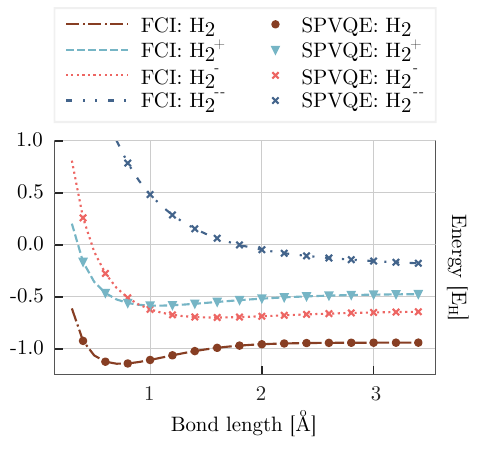}
    \caption{
    Energies of $\mathrm{H_2}$, $\mathrm{H_2^+}$, $\mathrm{H_2^-}$ and $\mathrm{H_2^{--}}$ as a function of the bond length.
    Markers represent noiseless SPVQE simulations, while lines represent exact FCI calculations in STO-6G basis for comparison.}
    \label{fig_H2}
\end{figure}

The dissociation profiles computed with SPVQE are presented in \Cref{fig_H2} and compared to the numerically exact solutions obtained with the classical Full Configuration Interaction (FCI) method \cite{Szabo1947}.

In all simulations presented in \Cref{fig_H2} SPVQE is able to target the physically correct state among the various ones that are present.

\begin{figure}[ht]
    \includegraphics[width=1\columnwidth]{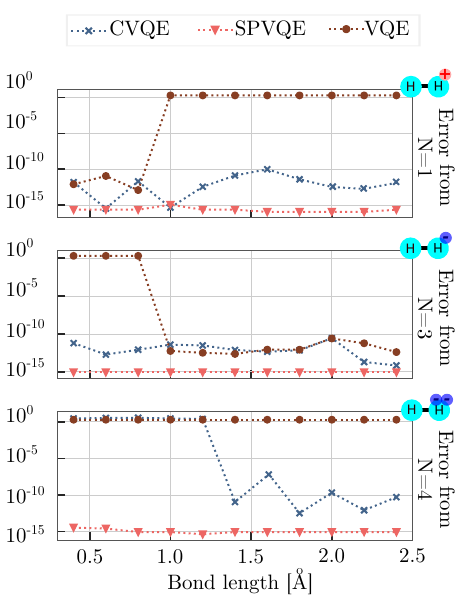}
    \caption{Analysis of $\mathrm{H_2^+}$, $\mathrm{H_2^-}$ and $\mathrm{H_2^{--}}$ (from the top to the bottom) for different bond lengths. Error on the correct value of $N$ computed among different noiseless simulations with random starting parameters using CVQE (best result among 100 simulations), SPVQE (best result among 2 simulations) and the standard VQE (best result among 1000 simulations).}
    \label{fig_error_spin_H2}
\end{figure}

Then, we repeated these simulations using CVQE and VQE and show a comparison of the different methods in \Cref{fig_error_spin_H2}. 
VQE points correspond to the best value obtained among 1000 simulations with random starting parameters. 
CVQE simulations were repeated 100 times and SPVQE calculations 2 times.

The standard VQE algorithm  struggles to find a good approximation for every configuration. 
In particular, it swaps $\mathrm{H_2^+}$ and $\mathrm{H_2^-}$ taking the state with the lowest energy between them (for that specific bond length) and is never able to approximate $\mathrm{H_2^{--}}$ that has a much higher energy.
CVQE reaches good results for both $\mathrm{H_2^+}$ and $\mathrm{H_2^-}$.
However, for $\mathrm{H_2^{--}}$ it needs a higher penalty (in particular at low bond lengths) and this misleads the algorithm that fails to find the correct minimum.
Lastly, SPVQE reaches the correct state at every configuration, hitting machine precision.

\subsection{Dependence on the number of steps}
\label{ssec:number_steps}

\begin{figure}
    \centering
    \includegraphics[width=1\columnwidth]{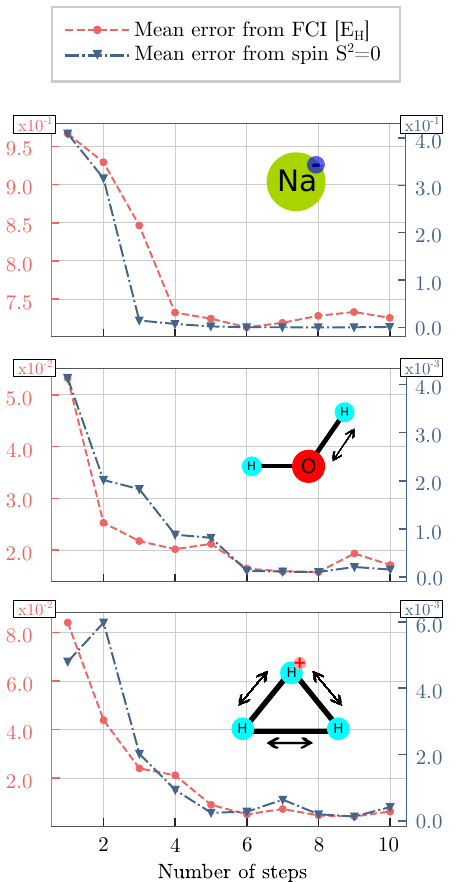}
    \caption{Results for $\mathrm{Na^-}$, $\mathrm{H_2O}$ and $\mathrm{H_3^+}$ with $ \mu_{max} = 1$ constraint on the total spin operator. 
    Every point corresponds to a full SPVQE calculation with a different number of steps.
    Increasing the number of sequence steps improves the accuracy of our variational approximation, both for the energy and the total spin, which is the constrained operator.
    }
    \label{Fig:iterations}
\end{figure}

In this section we analyze the performance of SPVQE as a function of the number of sequence steps $N_s$.
In \Cref{Fig:iterations}, we show the results for three different molecules: $\mathrm{Na^-}$, $\mathrm{H_2O}$ and $\mathrm{H_3^+}$. 

While $\mathrm{H_2O}$ is not an excited state, the addition of a constraint is still needed because the molecule has an excited level with a very similar energy, making the convergence of standard VQE to a state with the right physical properties difficult \cite{Kandala2017, Ryabinkin2019}.

For $\mathrm{H_2O}$ we varied the length of one of the $\mathrm{H-O}$ bond from $1\,\si \angstrom$ to $3\,\si \angstrom$, while for $\mathrm{H_3^+}$ we varied the length of all $\mathrm{H-H}$ bonds from $1\,\si \angstrom$ to $3\,\si \angstrom$.
These configurations are difficult to handle for the standard VQE, which is not able to reach a correct approximation.

For $\mathrm{Na^-}$ and $\mathrm{H_2O}$ the simulations were carried in the active space, freezing core orbitals to diminish the number of needed qubits.
This resulted in an ansatz with 24 parameters on 6 qubits for $\mathrm{Na^-}$,considering 2 electrons in the 4 most external molecular orbitals, and one with 16 parameters on 4 qubits for $\mathrm{H_2O}$, considering just 4 electrons in 3 orbitals.
$\mathrm{H_3^+}$ simulations are performed without frozen core approximation, requiring 4 qubits and 16 parameters.
Thus, every marker on the graphs represents the average result obtained considering a SPVQE iteration for each ionic configuration.

The plots show that even with few iterations of the SPVQE method, the total spin is correctly constrained to the desired value, while the energy decreases accordingly.
We note that the first marker of each graph, namely the single-step optimization,is equal to a CVQE calculation.

\subsection{Robustness against starting point choice}
\label{ssec:robustness_start_par}

Finding the ground state of a quantum system is a hard problem even on quantum computers \cite{kitaev02_qc}.
For this reason, VQE and its modifications are not guaranteed to reach the global minimum of the energy (or the modified cost function) in polynomial time in every scenario.
The convergence of the algorithm depends on different factors.
The choice of starting parameters is one of them and can hinder the convergence of the algorithm.
Despite this importance, often we do not have the possibility to make an educated guess on the starting set of parameters; thus a certain robustness with respect to the choice of the starting parameters is desired.

\begin{figure}[h]
    \includegraphics[width=1\columnwidth]{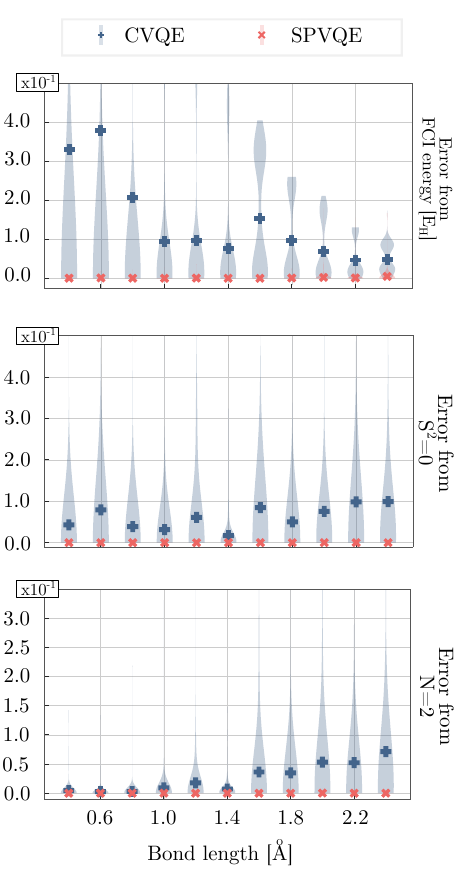}
    \caption{Energy, total spin and particle number error comparison between the variational approximation of the ground state of $\mathrm{H_3^+}$ at different bond length obtained with CVQE and SPVQE.
    The error is calculated as $|O_{\text{x-VQE}}-O_{\text{FCI}}|$, where $O$ is the observable considered.
    Every marker corresponds to the mean error of 100 different simulations with random starting parameters. Colored areas mirror the underlying data distribution.}
    \label{fig_starting_points}
\end{figure}

In this section we analyze the robustness of SPVQE and compare it to CVQE.
To this aim, we performed multiple simulations of the $\mathrm{H_3^+}$ molecule for bond lengths varying from $0.4\,\si\angstrom$ to $2.5\,\si\angstrom$. 
Results of these calculations are shown in \Cref{fig_starting_points}.
Every marker is obtained considering multiple simulations with random starting parameters and represents the mean error of the computed energies, numbers of particle, spins.
The coloured areas of the \textit{violin-plot}  mirror the underlying data distribution.

SPVQE proves to be more robust in every analyzed situation, with lower errors and smaller deviations from the mean value.
This is due to the fact that SPVQE is generating itself a sequence of improving educated guesses.

To show that this effect is not due to some inherent property of the $\mathrm{H_3^+}$ molecule, in \ref{app_robust_pars} we report the calculations performed on different chemical systems.

\subsection{Hardware experiments}

Finally, we  tested SPVQE on real quantum hardware, using the IBM Quantum processor \textit{ibmq\_quito}.

We analyzed the trihydrogen cation ($\mathrm{H_3^+}$) for bond lengths varying from $0.3\,\si\angstrom$ to $2.5\,\si \angstrom$.
A standard VQE approach works properly for bond lengths smaller than $\approx 1.5\,\si\angstrom$: after that point, the energy of $\mathrm{H_3^-}$ gets lower than the desired configuration, making the algorithm converge to the wrong minimum. 
Thus, the optimization process fails to preserve the particles number (it finds a state where $N=4$) and also the spin (it computes $S^2=1$).
Therefore we chose to constrain $S^2=0$.

As we stated before, current quantum hardware has limited performances and requires extreme care to reduce errors impact. 
Since \textit{ibmq\_quito} has a medium-low quantum volume \cite{Cross2019} of 16, we reduced the depth of the ansatz described in \Cref{sec:num_sim} choosing $D=2$.
Once the circuit is transpiled using \textit{ibmq\_quito} basis gates \{CX, ID, RZ, SX, X\}, its  final depth is 19.
For every bond length configuration we performed 4 independent calculations with 1024 shots each.
Starting parameters have been chosen randomly.

Hardware experiments were prepared simulating the SPVQE algorithm in a noisy environment. 
The obtained results, reported in \ref{app_noise}, confirmed us that SPVQE could mantain the desired advantages even in the presence of noise.

\begin{figure}[h]
    \includegraphics[width=1\columnwidth]{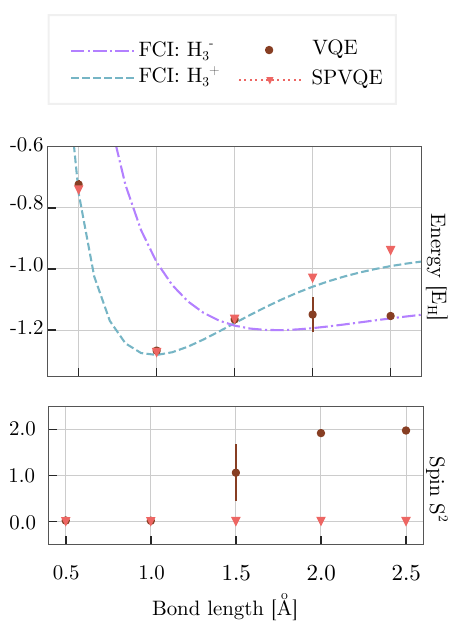}
    \caption{Energy and spin computed for SPVQE and VQE on quantum hardware. Dashed and dashdotted lines correspond to FCI calculations for the dissociation profile of $\mathrm{H_3^+}$ and $\mathrm{H_3^-}$.
    Error bars represent statistical errors computed over multiple experiments.}
    \label{fig_hardware}
\end{figure}

Hardware results are showed in \Cref{fig_hardware}.

We can see that the profile computed using SPVQE correctly approximate the exact calculated one.
VQE fails to correctly approximate the state of $\mathrm{H_3^+}$ at high bond distance, when $\mathrm{H_3^-}$ energy level became lower than the $\mathrm{H_3^+}$ one.
Error bars correspond to the standard deviation obtained among the 4 different results and thus represent the statistical error.

VQE shows the biggest statistical uncertainty at at $2\,\si\angstrom$.
For this configuration, the VQE wave functions always present the wrong spin.
Therefore, this point can be interpreted as corresponding to a molecular configuration where it is difficult to find the global minimum of the Hamiltonian.

Looking at the lower panel of \Cref{fig_hardware}, we see that SPVQE is able to constrain $S^2$ to the desired value even on hardware, while VQE starts to miscalculate it at $1.5\,\si\angstrom$, where the energies of the two different spin configurations are similar.
Moreover, the distance where the energies of $\mathrm{H_3^+}$ and $\mathrm{H_3^-}$ are almost equal has the biggest statistical uncertainty on the spin.
This can be explained by the fact that VQE just minimizes the energy, thus finding a state not characterized by a precise spin, but rather by a mixed spin state.

\section{Discussion}
\label{sec:discussion}

In this paper we introduced a simple, yet effective, modification of the VQE algorithm to enable excited and ionized state simulations.
The proposed method, Sequence of Penalties VQE, was tested for small molecules both on classical simulators and on real quantum hardware and proved to work correctly in the studied cases.
Moreover, the algorithm showed an increased robustness against the choice of the starting point.
For these reasons, we envision to use SPVQE in combination with other methods to study excited states of physical systems with higher accuracy on quantum computers.

The choice of the parameters, number of iterations and maximum penalty is done heuristically and can be further optimized in future studies.
Then, another outlook is the integration with error mitigation techniques \cite{Giurgica_Tiron_2020,Temme2017,Pokharel2018,Kandala2019}, in order to improve the accuracy of the obtained results and scale the method to bigger physical systems.

In the end, we think that SPVQE should be taken in consideration for calculations regarding excited states, considering the straightforwardness of its implementation, its reliability and the accuracy improvements that it brought.

\appendix

\section{Computational details for frozen-core approximation}\label{app_details}

For all the simulations and computations, we used the Qiskit package \cite{Qiskit} (version 0.39.0 with qiskit-nature module at version 0.3.0) together with the PySCF package \cite{Sun2018} (version 2.1.1).

To reduce computational burden we restricted the active space with the Qiskit ActiveSpaceTransformer.
All the computed values, including frozen core energies and nuclear repulsion energies, are available on GitHub\cite{Github}.

In \Cref{tab:sim-details} we summarize some of the most important simulation details for every studied molecule.
Here, bold symbols correspond to computations on hardware.

\begin{table*}
\centering
\rowcolors{1}{}{lightgray}
\begin{tabular}{l|ccccc}
\multicolumn{1}{c|}{} & Active electrons & Active orbitals & Qubits & D ansatz & \multicolumn{1}{c}{Number of  parameters}  \\ \hline
$\mathrm{H_2}$   & 2 (all) & 2 (all) & 2 & 3 & 8  \\
$\mathrm{H_3^+}$ & 2 (all) & 3 (all) & 4 & 3 & 16  \\
$\boldsymbol{\mathrm{H_3^+}}$ & 2 (all) & 3 (all) & 4 & 2 & 12 \\
$\mathrm{Na-}$   & 2 & 4 & 6 & 3 & 24 \\
$\mathrm{H_2O}$  & 4 & 3 & 4 & 3 & 16 \\
$\mathrm{N_2}$   & 6 & 3 & 4 & 3 & 16 
\end{tabular}
\caption{Computational details for the molecules simulated in this paper. The bold $\boldsymbol{\mathrm{H_3^+}}$ corresponds to computations on hardware, where a shallower ansatz was used. }
\label{tab:sim-details}
\end{table*}

\section{Improving constrained calculations}
\label{app_post_processing}

To improve both CVQE and SPVQE results in noisy and real quantum environments, we compute the expectation value of the Hamiltonian one last time after the algorithm has converged. 
In fact, even if we have the value of the cost function $F$ and the penalty $ P$ correspondent to the best result found, the introduction of a penalty leads to a larger uncertainty. \\
We can show that with error propagation theory:

\begin{equation}
    E=F- P\rightarrow \sigma_E = \sqrt{\sigma^2_F + \sigma^2_P}
\end{equation}

Measuring a last time the expectation value of the Hamiltonian, using the best computed parameters, give us a measure of the best energy with the error associated minimized, because we have:

\begin{equation}
    E=F \rightarrow \sigma_E = \sigma_F < \sqrt{\sigma^2_F + \sigma^2_P}
\end{equation}

\section{Robustness of SPVQE for different molecular systems}\label{app_robust_pars}

While in \Cref{ssec:robustness_start_par} we presented the result for $\mathrm{H_3^+}$, we now show the same analysis for different molecules.
This analysis shows that the robustness of SPVQE is not limited to the $\mathrm{H_3^+}$ system.
We simulated, in a noiseless environment, different molecules applying SPVQE and CVQE multiple times with different and random starting parameters.

\begin{figure}[htp]
    \includegraphics[width=1\columnwidth]{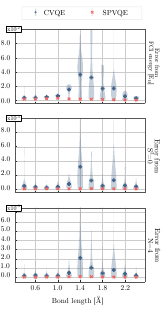}
    \caption{
    Energy, total spin and particle number error comparison between the variational approximation of the ground state of $\mathrm{H_2O}$ at different bond length obtained with CVQE and SPVQE.
    Every marker corresponds to the mean error of 100 different simulations with random starting parameters.}
    \label{fig_starting_points_h2O}
\end{figure}

\begin{figure}[htp]
    \includegraphics[width=1\columnwidth]{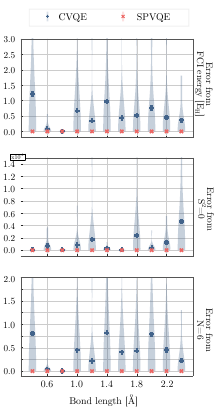}
    \caption{Same as \Cref{fig_starting_points_h2O} for $\mathrm{N_2}$. }
    \label{fig_starting_points_n2}
\end{figure}

In particular, we studied the $\mathrm{H_2O}$ molecule for different bond lengths (we varied the position of one hydrogen atom in respect to $\mathrm{H-O}$. To simulate a molecule with 10 electrons, we froze the core orbitals, leaving just 4 particles to be placed in 3 molecular orbitals.

We highlight that results both of CVQE and SPVQE were obtained setting the same number of iteration for the optimizer.

We also repeated the same analysis for the $\mathrm{N_2}$ molecule where we froze the core orbitals considering only 6 electrons in 3 molecular orbitals. 

Results are shown in \Cref{fig_starting_points_h2O,fig_starting_points_n2}.

\section{Noise simulations of SPVQE}
\label{app_noise}

We simulated $\mathrm{H_3^+}$ dissociation profile considering a noise model imported from the \textit{ibm\_perth} IBM Quantum processor.
Trihydrogen cation has 6 spin orbitals on the lowest shell: we therefore need a minimum of four qubits and a 16 parameters ansatz.
For every method, to help the optimization, we used as starting ansatz parameters the optimal ones calculated for the last computation of that same method. 

\begin{figure}[H]
    \includegraphics[width=0.95\columnwidth]{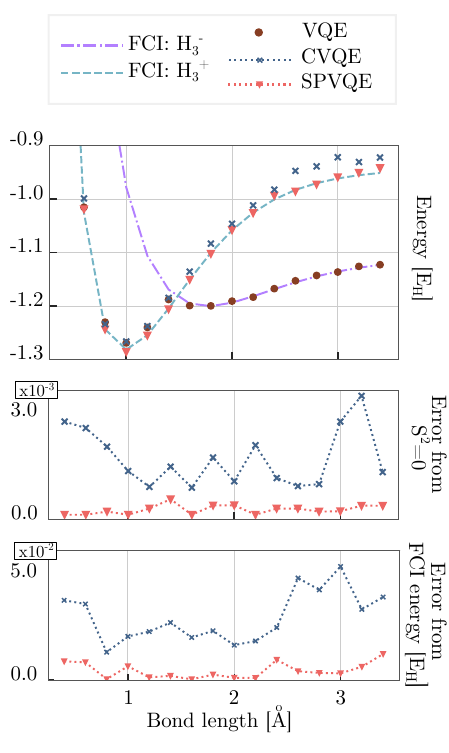}
    \caption{Energy calculated with VQE, CVQE and SPVQE for different bond lengths of the $\mathrm{H_3^+}$ molecule in a noisy simulation, together with error on FCI energy and total spin.
    The dashed and dashdotted lines represent FCI calculations for $\mathrm{H_3^+}$ and $\mathrm{H_3^-}$.}
    \label{fig_H3+}
\end{figure}

The results are presented in \Cref{fig_H3+}.

We can see that SPVQE outperforms both the standard VQE and the CVQE approach.
The VQE fails to converge to the right physical state for bond lengths greater than $1.4 \si\angstrom$.

Compared to CVQE, the SPVQE approach converges to the desired value with higher precision both for total spin (the constrained operator) and energy.

\end{document}